\documentclass[aps,showpacs,nofootinbib,superscriptaddress]{revtex4}
\usepackage{graphicx}

\begin{document}
\title{Two Particle-Hole Excitations in Charged Current Quasielastic Antineutrino--Nucleus Scattering}

\author{J. Nieves}
\affiliation{Instituto de F\'\i sica Corpuscular (IFIC), Centro Mixto
Universidad de Valencia-CSIC, Institutos de Investigaci\'on de
Paterna, E-46071 Valencia, Spain}
\author{I. \surname{Ruiz Simo}}
\affiliation{Dipartimento di Fisica, Universit\`a di Trento, I-38123 Trento, Italy}
\author{M. J.  \surname{Vicente Vacas}}
\affiliation{Departamento de F\'\i sica Te\'orica and IFIC, Centro Mixto
Universidad de Valencia-CSIC, Institutos de Investigaci\'on de
Paterna, E-46071 Valencia, Spain}

\today

\begin{abstract}
We evaluate the quasielastic and multinucleon contributions to the antineutrino nucleus scattering cross section and compare our results with the recent MiniBooNE data. We use a local Fermi gas model that includes RPA correlations and gets the multinucleon part from a systematic many body expansion of the $W$ boson selfenergy in the nuclear medium. 
 The same model had been quite successful for the neutrino cross section and contains no new parameters.
We have also analysed the relevance of 2p2h events for the antineutrino energy reconstruction.
\end{abstract}


\maketitle

Neutrino (antineutrino) charged-current (CC) scattering on nuclei without emitted pions is a crucial detection channel for many  neutrino experiments~\cite{Gran:2006jn,Lyubushkin:2008pe,AguilarArevalo:2010zc} and has been extensively studied by several theoretical groups~\cite{Benhar:2005dj,Nieves:2004wx,Nieves:2005rq,Leitner:2006ww,Martini:2009uj,Martini:2010ex,Leitner:2010kp,Benhar:2010nx,
Amaro:2010sd,Amaro:2011aa,Amaro:2011qb,Juszczak:2010ve,Bodek:2011ps,Butkevich:2010cr,Nieves:2011pp,Nieves:2011yp,Nieves:2012yz,Meucci:2012yq}.
It has been customary to take for granted that most of those events\footnote{After subtraction of some background, such as pion production followed by absorption.}
could be attributed to the quasielastic (QE) scattering of the weak probe on a nucleon. Hence, the initial  neutrino(antineutrino) energy could be approximately determined from the energy and angle of the final lepton  assuming QE kinematics.

This assumption soon led to some puzzling results. In particular, neutrino MiniBooNE data~\cite{AguilarArevalo:2010zc} were quite challenging showing substantially larger cross sections than the theoretical predictions~\cite{Boyd:2009zz,AlvarezRuso:2010ia}. A simple way out was to modify in the calculations
some standard hadronic parameters, such as the nucleon axial mass $M_A$~\cite{AguilarArevalo:2010zc,Butkevich:2010cr,Benhar:2010nx}, to fit data. However, the large obtained  values (e.g. $M_A=1.35\pm 0.17$ GeV in \cite{AguilarArevalo:2010zc}) were in clear  conflict with other electron and even neutrino experiments which result in a value very close to one ($M_A=1.014\pm 0.014$ GeV in~\cite{Bodek:2007ym}).

A more natural solution came from the incorporation of some standard nuclear effects such as RPA and multinucleon mechanisms. The importance of these effects  in the non mesonic neutrino nucleus scattering was first explored in Refs.~\cite{Martini:2009uj,Martini:2010ex} and later in 
Refs.~\cite{Amaro:2010sd,Nieves:2011pp,Nieves:2011yp}. Some of these more complete models were found to describe well even the MiniBooNE double differential cross section while using standard values for all parameters of the calculations~\cite{Nieves:2011yp}. In these models, a substantial part of the cross section measured in Ref.~\cite{AguilarArevalo:2010zc} corresponds to events in which at least two nucleons are emitted.

The relevance of the multinucleon mechanisms  has some unwanted consequences. Obviously, the neutrino energy reconstruction, based on the QE kinematics is not so reliable~\cite{Nieves:2012yz,Martini:2012fa,Lalakulich:2012hs,Martini:2012uc} and that implies larger systematic 
uncertainties in the experiments analysis. Furthermore, a clear separation of quasielastic  and multinucleon events is not straightforward from 
the experimental point of view, even if  final nucleons were detected, due to the energy thresholds and to the strong final state interaction of the nucleons in their way out of the nucleus. It is evident the need for comprehensive and theoretically well founded models that provide a 
good description for single and multinucleon processes.  Only with their help, one could approach the tasks of disentangling the genuine QE events and properly reconstruct the neutrino energy.
Of particular interest are the atomic number and the neutrino energy dependence of the reactions, the quest for kinematical regions with minimal multinucleon contamination and the relative importance of multinucleon mechanisms for neutrino versus antineutrino  induced reactions. In fact, the latter observable could be an stringent test for the available models~\cite{Bodek:2011ps,Lalakulich:2012ac}. Furthermore, if the contribution of multinucleon mechanisms is substantially different in neutrinos and antineutrinos, as predicted for instance in Refs.~\cite{Martini:2010ex,Amaro:2011aa} and this is not properly understood, it could lead to an asymmetry between $\nu$ and $\bar{\nu}$ which could be misinterpreted as a consequence of CP violation.

Our approach for the study of QE and multinucleon processes induced by neutrinos was developed in Refs.~\cite{Nieves:2005rq,Nieves:2011pp,Nieves:2011yp,Nieves:2012yz, Nieves:2004wx}. The model starts from a  local Fermi gas 
(LFG) picture for the nucleus. The purely QE contribution for both neutrinos and antineutrinos was studied in 
Refs.~\cite{Nieves:2004wx,Nieves:2005rq} incorporating among other nuclear effects Pauli blocking, Fermi motion and
the long range correlations (RPA). A correct energy balance is imposed using the experimental $Q$ values. Of the various options discussed in those papers, we will use here the fully relativistic model of 
Ref.~\cite{Nieves:2004wx}. Multinucleon mechanisms (not properly QE) in the neutrino nucleus scattering were studied in 
Ref.~\cite{Nieves:2011pp}. The approach is based on a systematic many body expansion of the $W$ boson absorption modes that includes one, two and even three nucleon mechanisms, as well as the excitation of $\Delta$ isobars followed by their absorption and the production of pions. The parameters of the calculation had
been previously fixed in the study of photon, electron and pion interactions with nuclei. The model successfully describes~\cite{Nieves:2011yp} the published MiniBooNE neutrino CCQE-like data, finding  a large contribution of two nucleon excitation mechanisms or, using the usual many body naming, 2 particle -- 2 hole (2p2h) channels. In a latter work~\cite{Nieves:2012yz}, we investigated the energy reconstruction procedure with the aim of determining the  influence of the multinucleon excitations on the  extraction of neutrino energy unfolded cross sections from the measured flux-average data. We found that the 2p2h contribution strongly distorts the shape of the total CCQE-like flux unfolded cross section for neutrino beams.

In this work, we apply the same model~\cite{Nieves:2011yp} for QE plus multinucleon mechanisms to antineutrino nucleus scattering and compare with the recent MiniBooNE data~\cite{MB2013}. The differences with the neutrino calculation are straightforward. The sign of the antisymmetric piece of the leptonic tensor  is changed and that implies a change on the sign of the parity violating pieces. In terms of the nuclear structure functions, this means a change of sign in the pieces proportional to $W_3$ in the differential cross section, following the notation of  Ref.~\cite{Nieves:2004wx}. The $W^+$ boson selfenergy in the nuclear medium is replaced by that of the $W^-$. This is achieved by  exchanging the role of protons and neutrons in all formulae. Finally, proper care has to be taken with the energy balance because of the different $Q$ values. We will always use
 the same axial mass of  our previous papers ($M_A = 1.049$ GeV).

The results presented in this paper correspond to QE and 2p2h contributions to the $\bar{\nu}_\mu+^{12}C\rightarrow \mu^++X$ process.  Scattering on other nuclei like $^{16}$O or $^{56}$Fe produce similar cross sections per nucleon.  Apart from 
the $\bar{\nu}$, $\nu$ energy dependence of the cross sections and their ratio and the muon angular distribution, we will also show  results  of the  double differential cross section $d^2\sigma /d cos\, \theta_\mu\, dT_\mu$ folded with the MiniBooNE $\bar{\nu}_\mu$ flux. 

\begin{figure}[htb]
\includegraphics[width=0.92\textwidth]{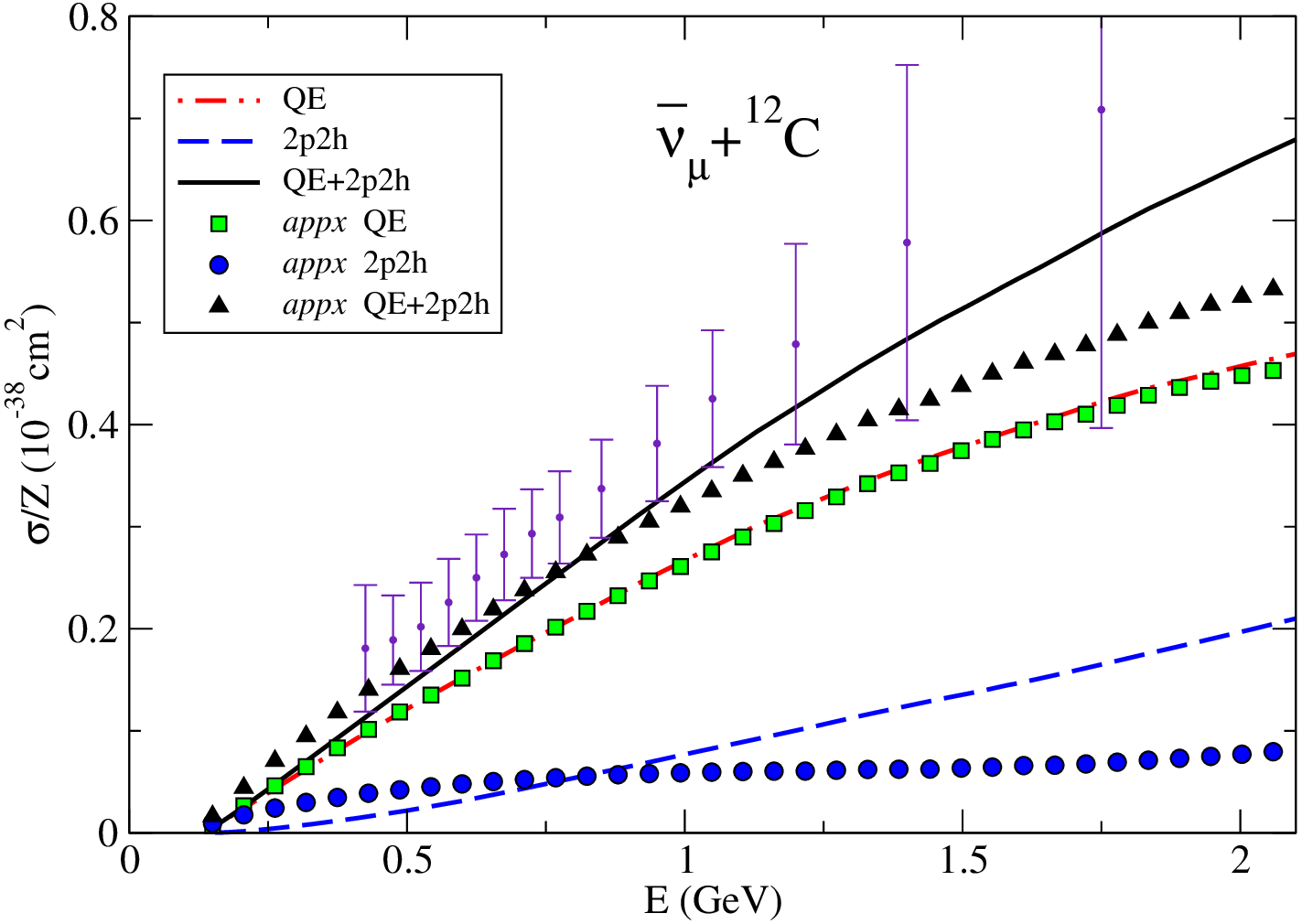}
\caption{Cross section contributions divided by the number of protons for $\bar{\nu}+^{12}C$ scattering as a function of the antineutrino energy. Curves labeled $appx$ correspond to the approximated unfolded cross section for the MiniBooNE flux. Experimental points from Ref.~\cite{MB2013}. }
\label{fig:1}
\end{figure}

The results are shown in Figs. 1-4.  In Fig.~\ref{fig:1}, we show the energy dependence of the QE and the 2p2h contributions to the   $\bar{\nu}_\mu+^{12}C$ cross section. The cross sections are smaller than for neutrinos. In this case, 2p2h channels also provide  an important contribution which cannot be neglected.  One of the relevant questions is to assess the quality of the standard procedures of energy reconstruction. 
Our results are similar to those obtained for the neutrino case. In the figure, we have labeled $appx$ the approximated unfolded cross sections\footnote{See Ref.~\cite{Nieves:2012yz} for details on the unfolding procedure}.  We observe that for genuine QE events the curve is hardly modified. However, for 2p2h events there is a substantial displacement towards low energies that is also reflected in the summed cross section. In other words, some ``high'' energy antineutrinos would be identified as ``low'' energy ones.
Although this statement is of a general validity, the detailed distortion of the cross section curve depends on the antineutrino flux.
The full model for the unfolded $\sigma_{appx}$ agrees well with the experimental data~\footnote{For this observable, there is an additional 17.2\% normalization uncertainty not shown in the figure.}, whereas the pure QE contribution predicts lower values for the cross section.

\begin{figure}[htb]
\includegraphics[width=\textwidth]{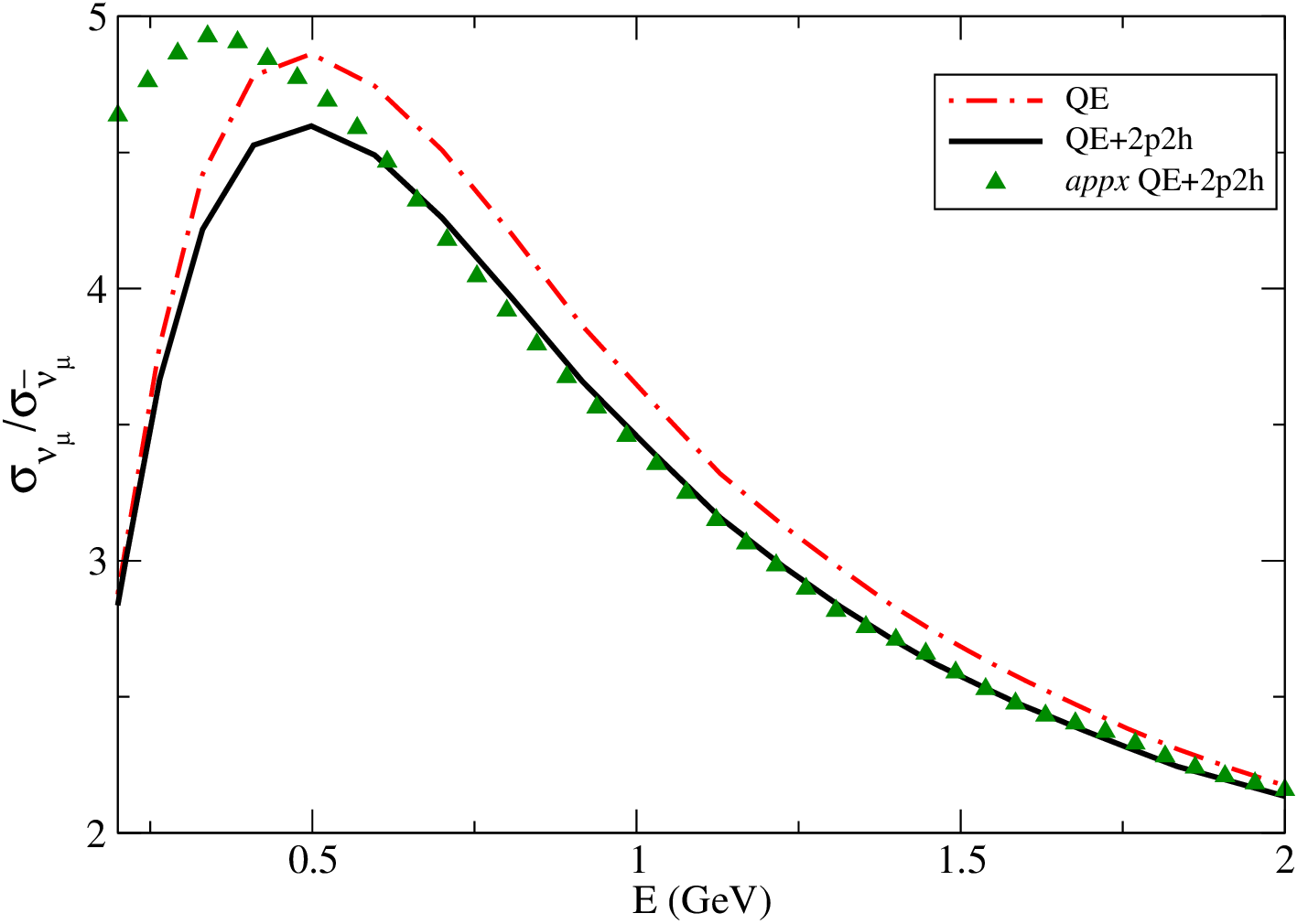}
\caption{Ratios of neutrino versus antineutrino cross sections in $^{12}C$ as a function of the energy. 
Solid line: QE + 2p2h. Dash-dotted line: QE. Triangles: Ratio obtained with the MiniBooNE flux using the unfolding procedure of Ref.~\cite{Nieves:2012yz}.}
\label{fig:2}
\end{figure}
Fig.~\ref{fig:2}, shows the ratio between the neutrino and the antineutrino induced reactions. The inclusion of the 2p2h mechanisms  slightly reduces the ratio (by less than a ten percent at the peak) without much affecting the shape.  Thus, in our model the relative importance of the 2p2h channel is somehow larger for antineutrinos.
A similar trend, although with a stronger reduction, has been found by Amaro et al.~\cite{Amaro:2011aa} in the superscaling approximation and by Meucci et al.~\cite{Meucci:2012yq} in the EDAI option of the relativistic Green's function model, which agrees well with the $\nu_\mu$ MiniBooNE data. Other works like Ref.~\cite{Bodek:2011ps}, which reaches agreement with MiniBooNE QE neutrino data by modifying the magnetic form factors of the bound nucleons, and Ref.~\cite{Martini:2010ex} lead to larger values for this ratio. This latter reference is of particular interest as it is the closest to our approach and contains a microscopical evaluation of 2p2h mechanisms similar to ours. However, there are some important differences which amount to a more comprehensive inclusion of mechanisms in our scheme and some approximations used in the calculations. A detailed discussion on these differences can be found in 
Ref.~\cite{Nieves:2011pp}. 
Finally, we find that the unfolding procedure respects the  $\sigma_\nu/\sigma_{\bar{\nu}}$  ratio for energies above 0.6 GeV but produces a strong distortion at lower energies.

\begin{figure}[htb]
\includegraphics[width=\textwidth]{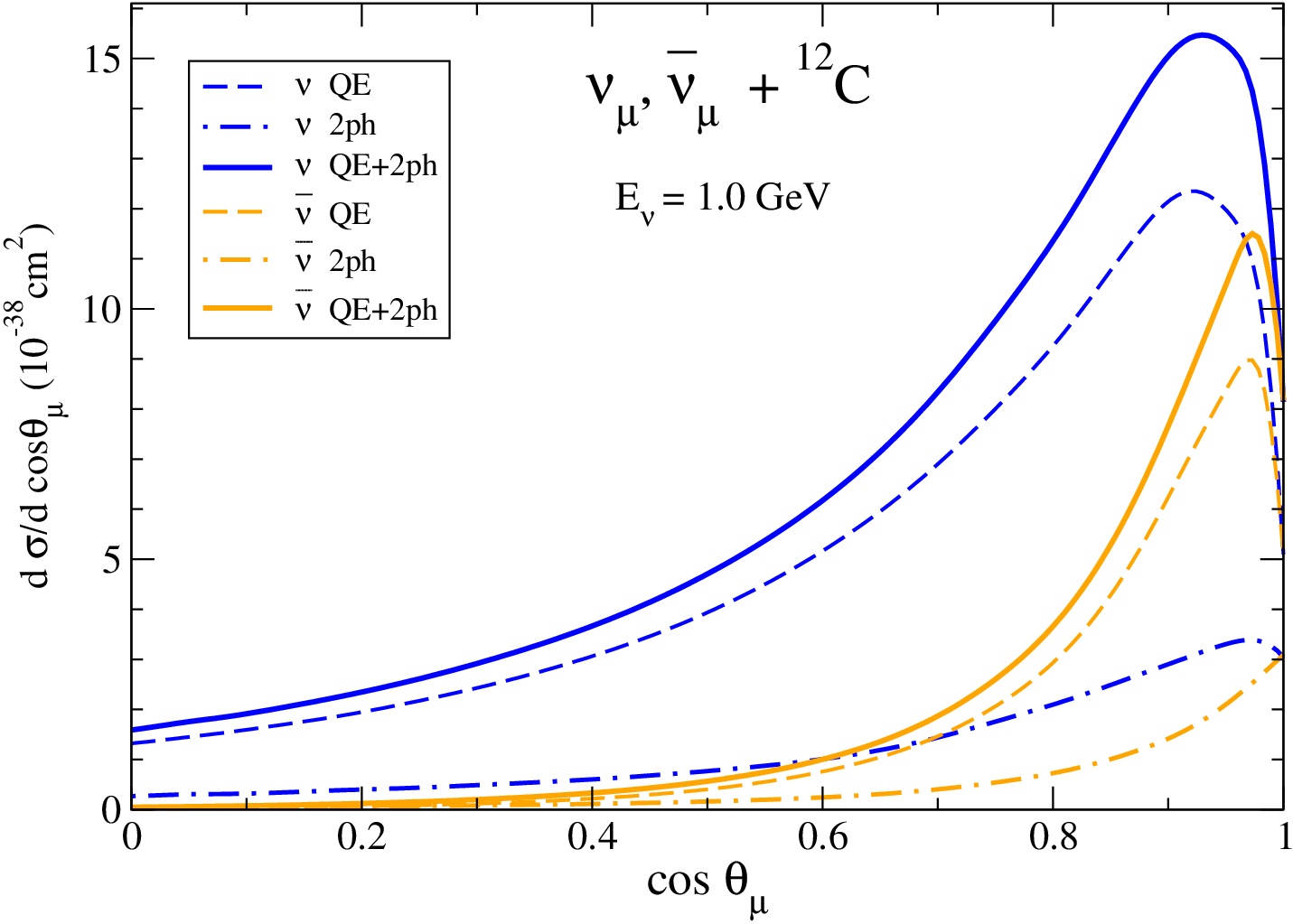}
\caption{Muon angular distributions for 1  GeV neutrinos (antineutrinos) scattering on $^{12}$C.}
\label{fig:3}
\end{figure}

In Fig.~\ref{fig:3}, we present  the muon angular distribution for 1 GeV neutrinos (antineutrinos). The $\bar{\nu}$ scattering is more forward peaked than the ${\nu}$ one  for both the QE and the 2p2h contributions. This is due to the different sign of the piece proportional to the $W_3$ structure function (cf. Eq. (10) of Ref.~\cite{Nieves:2004wx}) which leads to a negative interference for the $\bar{\nu}$ case. 

\begin{figure}[htb]
\includegraphics[width=\textwidth]{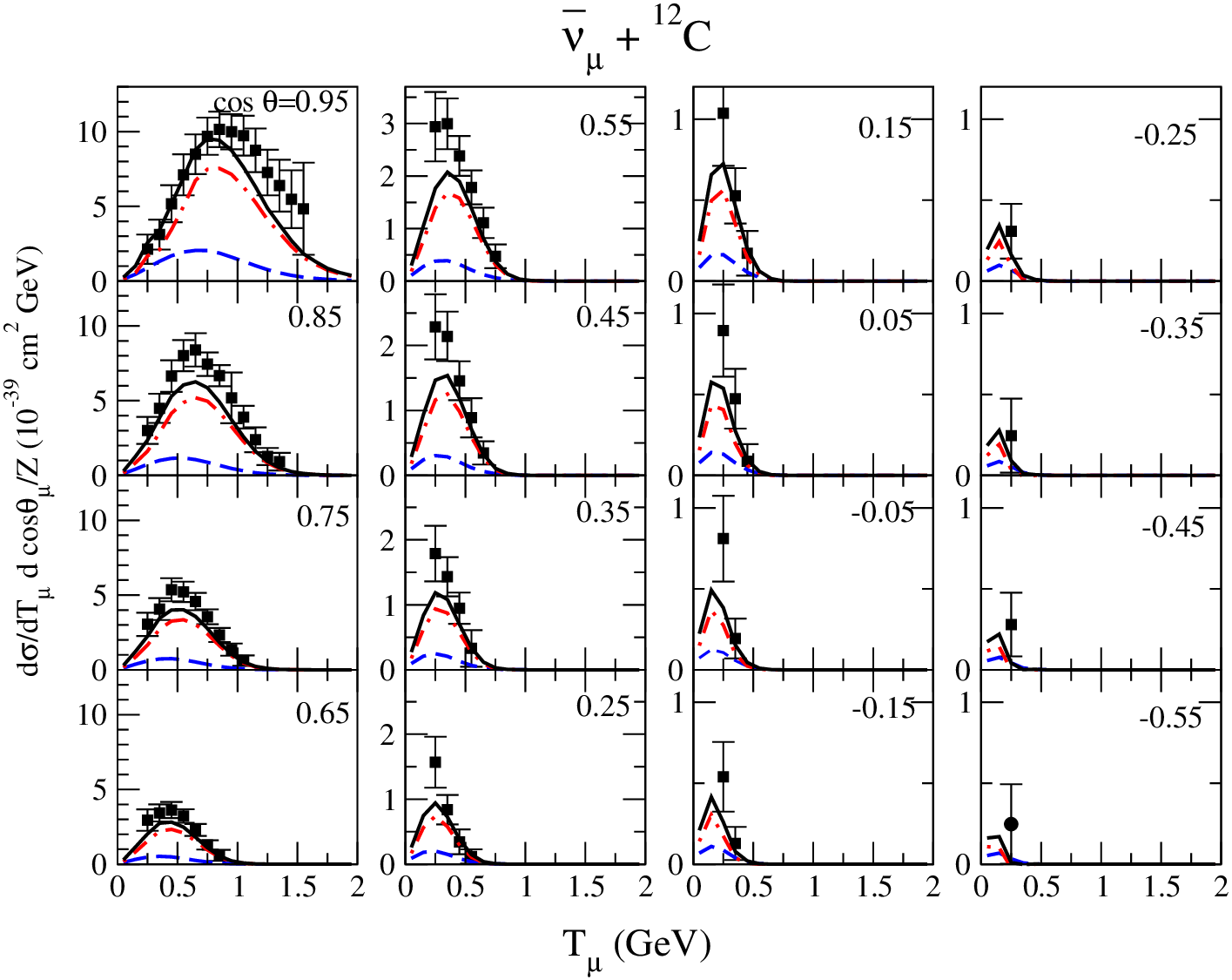}
\caption{Muon angle and energy distribution $d^2\sigma /d cos \theta_\mu dT_\mu$ per proton on a $^{12}$C target folded with
the MiniBooNE muon antineutrino flux. Different panels correspond to the various angular bins labeled by their cosinus central value. Black-solid line  is the full model including QE 
(relativistic and with RPA) and 2p2h mechanisms. Red-dash-dotted curve corresponds to QE and blue-dashed curve to 2p2h events. 
Data from Ref.~\cite{MB2013}, errors only account for the shape uncertainty. Besides, there is an additional normalization uncertainty of 17.2\%.}
\label{fig:4}
\end{figure}

As discussed in Ref.~\cite{Nieves:2011yp}, excluded the observation of final hadrons, a good observable to compare with theoretical models is the double differential cross section corresponding to the energy and angle of the final muon, because it avoids the problems of energy reconstruction and depends just on directly measured quantities.
In Fig.~\ref{fig:4}, we show our results for that cross section for the muonic antineutrino flux of MiniBooNE. 
In general terms, we successfully describe these data. For all angular windows 2p2h mechanisms play an important role and lead to a much better agreement with data. Their contribution tends to concentrate at low muon energies so that more energy is transferred to the nucleus than in  pure QE events.
Comparing this plot with its neutrino counterpart~\cite{AguilarArevalo:2010zc,Nieves:2011yp}, we observe a quite different pattern because of the more forward peaked character of the antineutrino induced reaction. There are also some changes concerning the antimuon energy distribution that are mostly due to the lower average energies of the MiniBooNE $\bar{\nu}$ flux.

In summary, we have studied non mesonic antineutrino nucleus cross sections in a model  that starts from a relativistic local Fermi gas description of the nucleus, includes RPA correlations and multinucleon effects finding a good agreement with experimental data. The same model had been quite successful for the neutrino cross section and contains no free parameters. The effects of multinucleon events are seen to be larger, relative to the QE contribution, than for the neutrino case and considerably improve the agreement with experiment. We have also found that the size of the 2p2h cross section limits the reliability of the unfolding procedures based on the assumption of QE kinematics for the non mesonic events.

\begin{acknowledgments}
This research was supported by
the Spanish Ministerio de Econom\'\i a y Competitividad
and European FEDER funds under the contracts FIS2011-28853-C02-01 and  FIS2011-28853-C02-02,
by Generalitat Valenciana under contract PROMETEO/2009/0090 and by the EU Hadron-Physics2 project, grant agreement no. 227431.
\end{acknowledgments}


\end{document}